\documentclass[11pt,showpacs,preprintnumbers,amsmath,amssymb,prd,nofootinbib,superscriptaddress]{revtex4-2}

\usepackage{dcolumn}% Align table columns on decimal point
\usepackage{bm}% bold math
\usepackage{ifpdf}
\usepackage{hyperref}
\usepackage{xcolor,color,graphicx,graphics,physics}
\usepackage[spanish,english]{babel}%, portuguese
\usepackage[latin1]{inputenc}
\usepackage[OT1]{fontenc}
\usepackage{latexsym,amssymb,amsmath,amsfonts, slashed,cancel, simpler-wick}
\usepackage{makeidx}
\usepackage{epsfig,subfigure}
\usepackage{natbib}
\usepackage{epstopdf}
\usepackage{mathrsfs}
\usepackage{hyperref}%\usepackage[colorlinks=true,linkcolor=blue]{hyperref}
\hypersetup{colorlinks=true, linkcolor=blue, citecolor=blue, urlcolor=blue}
\usepackage{enumerate}

\usepackage{fixmath}

\everymath{\displaystyle}
\usepackage{graphicx}

\usepackage[T1]{fontenc}
\usepackage{amsmath}
\usepackage{amssymb}
\usepackage{graphicx}
\usepackage{xcolor}

\newcommand{\bea}{\begin{eqnarray}}
\newcommand{\eea}{\end{eqnarray}}

\newcommand{\orcid}[1]{\href{https://orcid.org/#1}{\includegraphics[width=10pt]{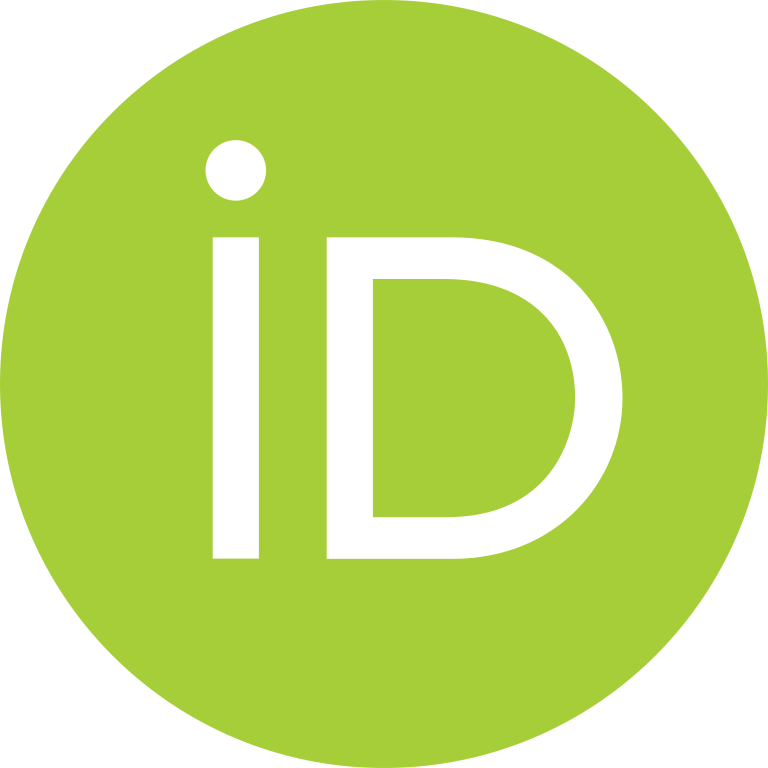}}}

\begin{document}

\title{Lorentz-violating Yukawa theory at finite temperature}

\author{D. S. Cabral  \orcid{0000-0002-7086-5582}}
\email{danielcabral@fisica.ufmt.br}
\affiliation{Programa de P\'{o}s-Gradua\c{c}\~{a}o em F\'{\i}sica, Instituto de F\'{\i}sica,\\ 
Universidade Federal de Mato Grosso, Cuiab\'{a}, Brasil}

\author{L. A. S. Evangelista \orcid{0009-0002-3136-2234}}
\email{lucassouza@fisica.ufmt.br}
\affiliation{Programa de P\'{o}s-Gradua\c{c}\~{a}o em F\'{\i}sica, Instituto de F\'{\i}sica,\\ 
Universidade Federal de Mato Grosso, Cuiab\'{a}, Brasil}

\author{J. C. R. de Souza \orcid{0000-0002-7684-9540}}
\email{jean.carlos@fisica.ufmt.br}
\affiliation{Programa de P\'{o}s-Gradua\c{c}\~{a}o em F\'{\i}sica, Instituto de F\'{\i}sica,\\ 
Universidade Federal de Mato Grosso, Cuiab\'{a}, Brasil}

\author{L. H. A. R. Ferreira \orcid{0000-0002-4384-2545}}
\email{luiz.ferreira@fisica.ufmt.br}
\affiliation{Programa de P\'{o}s-Gradua\c{c}\~{a}o em F\'{\i}sica, Instituto de F\'{\i}sica,\\ 
Universidade Federal de Mato Grosso, Cuiab\'{a}, Brasil}

\author{A. F. Santos \orcid{0000-0002-2505-5273}}
\email{alesandroferreira@fisica.ufmt.br}
\affiliation{Programa de P\'{o}s-Gradua\c{c}\~{a}o em F\'{\i}sica, Instituto de F\'{\i}sica,\\ 
Universidade Federal de Mato Grosso, Cuiab\'{a}, Brasil}

\begin{abstract}

This paper addresses Yukawa theory, focusing on the scattering between two identical fermions mediated by an intermediate scalar boson, considering the effects of thermal contributions and Lorentz symmetry breaking. Temperature is introduced into the theory through the TFD formalism, while Lorentz violation arises from a background tensor coupled to the kinetic part of the Klein-Gordon Lagrangian. Two important quantities are calculated: the cross-section for the scattering process and the modified Yukawa potential. The main results obtained in this work demonstrate that considering Lorentz symmetry breaking has several implications for changes in symmetries and physical states, while the presence of temperature is strongly related to the strength of the interaction. This interplay between symmetry breaking and temperature effects provides deeper insights into the behavior of the Yukawa theory under different conditions.

\end{abstract}

\maketitle

\section{Introduction}

Based on Fermi's studies of the interactions between protons and neutrons within the nucleus \cite{fermi}, Hideki Yukawa, in 1934, proposed that this reaction would occur in a manner similar to, but distinct from, electromagnetic interaction, thereby describing the nature of this scattering \cite{yukawa}. He developed the concept that the force between these particles can be described by what is now known as the Yukawa potential, which acts as a screened Coulomb potential - an electric interaction that is exponentially damped \cite{screened1, screened2}. This theory was highly successful, as it was the first prototype for the nuclear force and gave rise to meson theory \cite{nambuyukawa}. As a result, numerous papers have explored Yukawa theory across various contexts \cite{refyuk1, refyuk2, refyuk3, refyuk4}, including its application in gravitational studies \cite{yukawagravity}. Here, we refer to Yukawa theory as the body of knowledge derived from scattering processes that give rise to a Yukawa potential. In terms of quantum field theory, the process that gives rise to this famous and important quantity is the scattering of two fermions via a scalar boson \cite{peskin, ryder}. For simplicity, the reaction considered is $e^{-}e^{-} \to e^{-}e^{-}$ through a scalar boson $\phi$.

Colladay and Kostelecky, in their exploration of CPT and Lorentz violation (LV) within the Standard Model, examined the impact of these violations on the Yukawa sector in the context of the Standard Model Extension (SME). Their work aimed to derive constraints relevant to a unified theory of quantum mechanics and gravity \cite{kostelecky1, kostelecky2, kosteleckycs, kostelecky2004gravity}. Based on these investigations, several works address Lorentz violation from different perspectives, such as scattering processes \cite{scatter1, scatter3, cabral2023violation, santos2020gravitational}, the Casimir effect \cite{santos2019lorentz, ferreira2022tfd, santos2022corrections}, modified gravity \cite{jesus2020godel, santos2015godel, jesus2019ricci}, and many other intriguing subjects \cite{aguirrelorentz, araujo2021thermodynamic, aa2021lorentz, filho2021thermal, araujo2021higher}. Since the process that gives rise to the Yukawa potential involves a scalar particle, some studies have specifically focused on Lorentz symmetry breakdown in the Lagrangian of the boson sector \cite{altschul}. Additionally, there are other works that introduce Lorentz-violating terms alongside the Yukawa-type coupling, as seen in \cite{mmferreira}.

Another important aspect of quantum field theory that has garnered significant interest is the incorporation of thermal effects into scattering processes, using both real-time and imaginary-time formalisms \cite{temp00, temp000, khannatfd, temp1, temp2, temp3, temp4}. As an application, one real-time approach is the TFD formalism, which is commonly used to derive important quantities in quantum scattering, such as cross-sections and decay rates at tree level \cite{scatter2, santos2022temperature, santos2016quantized,santos2019thermal}. This approach requires two key assumptions: the doubling of the Hilbert space and the use of the Bogoliubov transformation. As a result, an important outcome is obtained for the propagator of any theory: it is divided into two parts: the standard propagator at zero temperature and a thermal component. Therefore, since both symmetry breakings and the influence of temperature on systems are inherent aspects of nature, it is natural to explore the effects of these features within Yukawa theory, which describes the non-relativistic interaction between nucleons and serves as a prototype for nuclear force. For these reasons, this paper focuses on examining the modifications and consequences of extensions due to Lorentz violation and thermal effects on the scattering cross-section and potential, both derived from Yukawa theory. 

This paper is organized as following. Section \ref{sec1} provides the necessary framework for describing the process and calculating the relevant quantities. It begins with an overview of the TFD formalism in Section \ref{sectfd} and continues with the derivation of the Lorentz-violating thermal propagator for scalar bosons in Section \ref{secpropagator}. Section \ref{secscattering} covers the Yukawa scattering, starting with the transition probability derived from Feynman diagrams in Section \ref{secdiagram} and the derivation {of the differential cross-section in Section \ref{seccross}}, including specific cases and special limits. Section \ref{secpotential} is dedicated exclusively to the Yukawa potential and the changes it undergoes, with results and consequences presented. The concluding remarks are found in Section \ref{secconclusions}.

\section{Thermal Yukawa theory in a Lorentz-violating context}\label{sec1}

This section presents the Lorentz-violating Yukawa theory. Thermal effects are then introduced using the TFD formalism. Starting from the free scalar theory, we can incorporate a term that breaks Lorentz symmetries by adding a traceless and symmetric background field $b^{\mu\nu}$ to the free Klein-Gordon Lagrangian  \cite{kostelecky2,altschul}, as follows
\begin{eqnarray}
    \mathcal{L}_{\text{Free}}^{(\phi)}=\frac{1}{2}\left(g^{\mu\nu}+b^{\mu\nu}\right)\partial_\mu\phi\partial_\nu\phi-\frac{1}{2}m_{\phi}^2\phi^2,\label{eq23}
\end{eqnarray}
which leads to the following modified equation of motion
\begin{eqnarray}
    \left(g^{\mu\nu}+b^{\mu\nu}\right)\partial_\mu\partial_\nu\phi+m_\phi^2\phi=0.\label{eq03}
\end{eqnarray}
Therefore, the solutions for the scalar field can be expressed as a sum of plane waves, given by
\begin{eqnarray}
    \phi(x)=\int\frac{d^3k}{(2\pi)^3}\frac{1}{\sqrt{2k_0}}\left[a_k^{\dagger}e^{-ikx}+a_k e^{ikx}\right].\label{eq04}
    \end{eqnarray}
    
From Eq. (\ref{eq03}), the dispersion relation is obtained as follows
\begin{eqnarray}
    k^2+k\cdot b\cdot k=m_\phi^2\quad\quad \text{or}\quad\quad \eta^{\mu\nu}k_{\mu}k_{\nu}=m_{\phi}^2,
\end{eqnarray}
with $\eta^{\mu\nu}=g^{\mu\nu}+b^{\mu\nu}$. Additionally, another important feature is the modified stress-energy tensor for the Lagrangian (\ref{eq23}), which is given by
\begin{eqnarray}
    T^{\alpha\beta}=\partial^\alpha\phi\partial^\beta\phi-\frac{1}{2}g^{\alpha\beta}\partial_\mu\phi\partial^\mu\phi+\frac{1}{2}g^{\alpha\beta}m_\phi^2\phi^2 +b^{\alpha\mu}\partial_\mu\phi\partial^\beta\phi-\frac{1}{2}g^{\alpha\beta}b^{\mu\nu}\partial_\mu\phi\partial_\nu\phi.\label{eq26}
\end{eqnarray}
Note that $b_{\mu\nu}$ is a two-rank real constant tensor that selects certain preferred directions in space-time.

On the other hand, for a free Dirac spinor field, the Lagrangian is given by
\begin{eqnarray}
    \mathcal{L}_{\text{Free}}^{(\psi)}=\Bar{\psi}(i\gamma^\mu\partial_\mu-m)\psi,
\end{eqnarray}
where the solution to this equation is given by
\begin{eqnarray}
    \psi (x) = \int \frac{d^3 p}{(2 \pi)^3}\frac{1}{\sqrt{2p_0}}  \sum_{s}\left[ u^s_p b^\dagger_s(p) e^{-ipx} + v^s_p c_s(p) e^{ipx}\right].\label{eq25}
\end{eqnarray}

For a scattering process involving both fields (\ref{eq04}) and (\ref{eq25}), we can consider them in their asymptotic limits, incorporating a specific interaction Lagrangian. In this way, we start with the following total Lagrangian
\begin{eqnarray}
\label{1}
    \mathcal{L}=\Bar{\psi}\left(i\gamma^\mu\partial_\mu-m\right)\psi+\frac{1}{2}\partial_\mu\phi\partial^\mu\phi +\frac{1}{2}b^{\mu\nu}\partial_\mu\phi\partial_\nu\phi -\frac{1}{2}m_\phi^2\phi^2-g\Bar{\psi}\phi\psi,\label{eq01}
\end{eqnarray}
which represents the Lorentz-violating Yukawa theory, with $g$ being the coupling constant. This is analogous to QED theory, where fermions interact through an intermediate particle, but here the intermediate particle is a scalar boson rather than a spin-1 gauge boson. The similarities are quite evident and become even more pronounced when considering a massless particle, i.e., $m_\phi=0$. 

The equation of motion for the scalar field described by (\ref{eq01}) is
\begin{eqnarray}
    \square\phi+m_{\phi}^2\phi=-g\bar{\psi}\phi\psi-b^{\mu\nu}\partial_\mu\partial_\nu\phi.
\end{eqnarray}
This is analogous to the $\nu-$th component of the Maxwell equation, $\partial_\mu F^{\mu\nu}=j^\nu$. From this, we can see that the Lorentz-violating (LV) term plays a significant role, providing a correction to the ``scalar'' interaction current. Consequently, the Yukawa theory behaves like a scalar Klein-Gordon field, similar to how gauge QED operates within its gauge field framework.

Significant results also emerge when studying these characteristics at finite temperature. One of the most commonly used formalisms for introducing temperature is the real-time TFD formalism, which will be discussed in the next section.

\subsection{Thermo Field Dynamics (TFD) Formalism}\label{sectfd}

The construction of TFD arises from the duplication of the Hilbert space into the usual and tilde spaces \cite{temp000, khannatfd, scatter2, santos2022temperature, santos2016quantized,santos2019thermal}. In other words, the thermal Hilbert space can be written as \(\mathbb{S}_T = \mathbb{S} \otimes \tilde{\mathbb{S}}\), where \(\mathbb{S}\) represents the usual Hilbert space and \(\tilde{\mathbb{S}}\) represents the dual Hilbert space. For an arbitrary operator $A$, the doubled notation leads to
\begin{eqnarray}
    A^a=\begin{pmatrix}
    A\\\chi\tilde{A}^\dagger
    \end{pmatrix},
\end{eqnarray}
where $\chi=1$ ($-1$) for bosons (fermions) and $a=1,2$. Another component of this construction is the Bogoliubov transformations, which connect the thermal and zero-temperature operators. They are given by $\phi^a=\mathbb{M}^{ab}\phi_b(\beta)$ and $\psi^a=\mathbb{N}^{ab}\psi_b(\beta)$, where the transformation matrices  $\mathbb{M}^{ab}$ and $\mathbb{N}^{ab}$  are defined for bosons and fermions, respectively, as
\begin{eqnarray}
    \mathbb{M}^{ab}(\beta,p_0)=\begin{pmatrix}
        U^\prime(\beta,p_0) & V^\prime(\beta,p_0) \\
        V^\prime(\beta,p_0) & U^\prime(\beta,p_0)
    \end{pmatrix},\quad\quad \mathbb{N}^{ab}(\beta,p_0)=\begin{pmatrix}
        U(\beta,p_0) & V(\beta,p_0) \\
        -V(\beta,p_0) & U(\beta,p_0)
    \end{pmatrix}.\label{eq20}
\end{eqnarray}
Here we have $(U^\prime)^2=1+n(\beta,p_0)$ and $(V^\prime)^2=n(\beta,p_0)$,  as well as $U^2=1-f(\beta,p_0)$ and $V^2=f(\beta,p_0)$, in addition to
\begin{eqnarray}
    n(\beta,p_0)=\frac{1}{e^{\beta p_0}-1},\quad\quad f(\beta,p_0)=\frac{1}{e^{\beta p_0}+1},
\end{eqnarray}
where these quantities represent the boson and fermion distribution functions, respectively. 

With these ingredients, a thermal vacuum state $\ket{0(\beta)}$ can be constructed. The thermal average of quantum operators $A$ can then be calculated using the following form
\begin{eqnarray}
    \langle A\rangle=\bra{0(\beta)}A\ket{0(\beta)}.
\end{eqnarray}

Now, let's briefly discuss the boson propagator and how to derive it within the TFD formalism, including the addition of the Lorentz violation term.

\subsection{Boson propagator at finite temperature with Lorentz violation}\label{secpropagator}

In the TFD formalism \cite{temp000, khannatfd}, the boson propagator is defined as
\begin{eqnarray}
    D_{F}^{ab}(x-y)&=&\bra{0(\beta)}\mathcal{T}\left[\phi^a(x)\phi^b(y)\right]\ket{0(\beta)}\nonumber\\
    &=&\Theta(x^0-y^0)\bra{0(\beta)}\phi^a(x)\phi^b(y)\ket{0(\beta)}+\Theta(y^0-x^0)\bra{0(\beta)}\phi^a(y)\phi^b(x)\ket{0(\beta)},\label{16}
\end{eqnarray}
where $\mathcal{T}$ is the time-ordering operator, $\Theta(x)$ is the step function and the $\phi^a(x)$ is the free scalar field operator (\ref{eq04}) in the doubled notation. Let's calculate each term separately. For the first term, we have
\begin{eqnarray}
\bra{0(\beta)}\phi^a(x)\phi^b(y)\ket{0(\beta)}&=&\int\frac{d^3qd^3k}{(2\pi)^3(2\pi)^3}\frac{1}{\sqrt{2q_0}\sqrt{2k_0}}\biggl[e^{-i q x}e^{ik y}\bra{0(\beta)}a^{\dagger a}(q)a^{b}(k)\ket{0(\beta)}\nonumber\\
&+&e^{i q x}e^{-ik y}\bra{0(\beta)}a^{a}(q)a^{\dagger b}(k)\ket{0(\beta)}\biggr].\label{eq18}
\end{eqnarray}
Using the Bogoliubov transformation we can write
\begin{eqnarray}
    \bra{0(\beta)}a^{\dagger a}(q)a^{b}(k)\ket{0(\beta)}&=&\mathbb{M}^{a2}(\beta,q)\mathbb{M}^{b2}(\beta,q)(2\pi)^3\delta^3(q-k)
\end{eqnarray}
and
\begin{eqnarray}
    \bra{0(\beta)}a^{a}(q)a^{\dagger b}(k)\ket{0(\beta)}=\mathbb{M}^{a1}(\beta,q)\mathbb{M}^{b1}(\beta,q)(2\pi)^3\delta^3(q-k).
\end{eqnarray}
Similar results are obtained for the second term of Eq. (\ref{16}). Thus, the boson propagator becomes
\begin{eqnarray}
    D^{ab}_F(x-y)&=&\int\frac{d^3q}{(2\pi)^3}\frac{1}{2q_0}\biggl[\Theta(y^0-x^0)\mathbb{M}^{a1}\mathbb{M}^{b1}+\Theta(x^0-y^0)\mathbb{M}^{a2}\mathbb{M}^{b2}\biggr]e^{-iq(x-y)}\nonumber\\
    &+&\int\frac{d^3q}{(2\pi)^3}\frac{1}{2q_0}\biggl[\Theta(x^0-y^0)\mathbb{M}^{a1}\mathbb{M}^{b1}+\Theta(y^0-x^0)\mathbb{M}^{a2}\mathbb{M}^{b2}\biggr]e^{iq(x-y)}.\label{eq02}
\end{eqnarray}

After performing some calculations and applying the residue theorem, we obtain
\begin{eqnarray}
    D_F^{ab}(x-y)=-i\int \frac{d^4q}{(2\pi)^4}\biggl[\frac{1}{q\cdot\eta\cdot q-m_{\phi}^2}\tau+2\pi i n\delta(q\cdot\eta\cdot q-m_{\phi}^2)\Delta(q_0)\biggr]e^{-iq(x-y)},
\end{eqnarray}
where the dispersion relation $(q_0\pm i\epsilon)^2\sim q_0^2\pm i\epsilon=m_\phi^2+\Vec{q}^2-q\cdot b\cdot q\pm i \epsilon$ has been used and 
\begin{eqnarray}
    \Delta(q_0)=\begin{pmatrix}
        1 & e^{\beta q_0/2}\\ e^{\beta q_0/2} & 1
    \end{pmatrix},\quad\quad\quad\quad \tau=\begin{pmatrix}
        1 & 0\\0 &-1
    \end{pmatrix}.
\end{eqnarray}
Additionally, a regularized delta function relation \cite{regdelta} has been considered, i.e.,
\begin{eqnarray}
    2\pi i\frac{1}{n!}\frac{\partial^n}{\partial x^n}\delta(x)=\left(-\frac{1}{x+i\epsilon}\right)^{n+1}-\left(-\frac{1}{x-i\epsilon}\right)^{n+1}.
\end{eqnarray}

On the other hand, knowing that, as a first approximation, we have the following
\begin{eqnarray}
    \frac{1}{\eta^{\mu\nu}q_{\mu}q_{\nu}-m_{\phi}^2-i\epsilon}-\frac{1}{\eta^{\mu\nu}q_{\mu}q_{\nu}-m_{\phi}^2+i\epsilon}&=&2\pi i \delta(q^2-m_\phi^2)-2\pi i b_{\mu\nu}q^\mu q^\nu\delta^{\prime}(q^2-m_{\phi}^2),\label{eq09}
\end{eqnarray}
the thermal propagator with Lorentz violation corrections can be written as
\begin{eqnarray}
     D_F^{ab}(x-y)&=&-i\int \frac{d^4q}{(2\pi)^4}\biggl\{\frac{1}{q\cdot\eta\cdot q-m_{\phi}^2}\tau\nonumber\\
     &+&2\pi i\left[\delta(q^2-m_\phi^2)- b_{\mu\nu}q^{\mu}q^{\nu} \delta^{\prime}(q^2-m_{\phi}^2)\right] n\Delta(q_0)\biggr\}e^{-iq(x-y)}.
\end{eqnarray}
Note that as the temperature approaches zero, the Lorentz-violating zero-temperature propagator, as obtained by \cite{altschul}, is recovered, that is,\begin{eqnarray}
    D_F(x)=-i\int\frac{d^4q}{(2\pi)^4}\frac{1}{\eta^{\mu\nu}q_{\mu}q_\nu-m_{\phi}^2}e^{-iqx}.
\end{eqnarray}

With these tools, the next section will explore the Yukawa scattering process with Lorentz violation at finite temperature.

\section{Yukawa scattering}\label{secscattering}

In this section, we will discuss Yukawa scattering with thermal effects and Lorentz violation symmetries. To calculate the cross section, we need to derive the average of the scattering amplitude and define a reference frame. For this study, we will use the Lab frame. Let's begin by determining the transition amplitude.

\subsection{The scattering}\label{secdiagram}

\begin{figure}[!htb]
    \centering
    \includegraphics[scale=0.3]{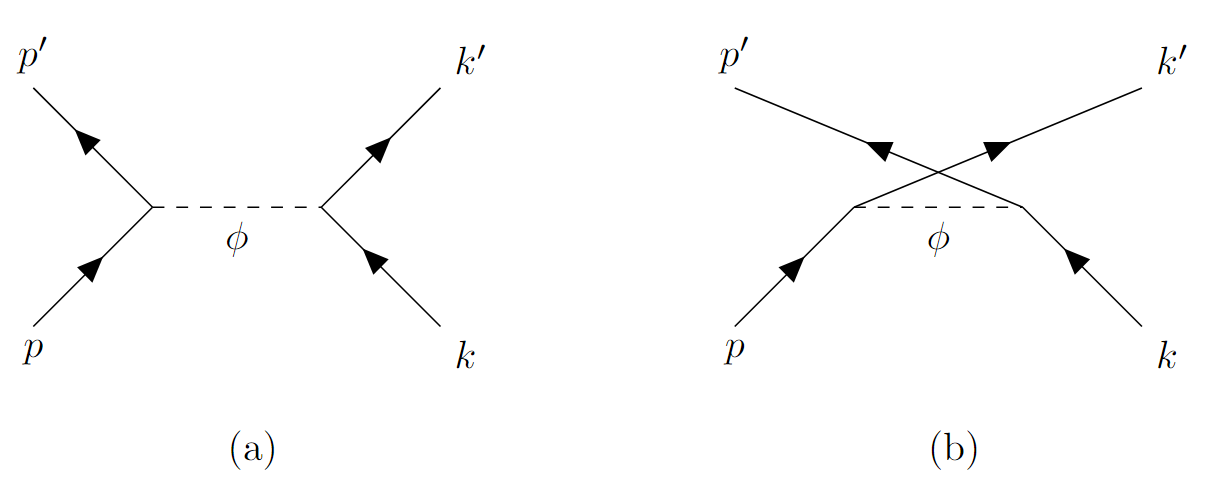}
    \caption{Feynman diagrams for fermion-fermion Yukawa scattering. Diagrams (a) and (b) represent the t-channel and u-channel, respectively.}
    \label{fig1}
\end{figure}

This subsection focuses on the interaction between fermions through Yukawa scattering at finite temperature. The Feynman diagrams illustrating this scattering process are shown in Figure \ref{fig1}.
In order to calculate the cross section, we need to determine the scattering amplitude $\mathcal{M}_{fi}^{(\beta)}$, which is given by
\begin{eqnarray}
    \mathcal{M}_{fi}^{(\beta)}=\bra{f(\beta)}\hat{S}^{(2)}\ket{i(\beta)},
\end{eqnarray}
where the thermal initial and final states are given as
\begin{eqnarray}
    \ket{i(\beta)} &=& b^{\dagger}_p(\beta) b^{\dagger}_k(\beta) \ket{0(\beta)} ,\\
    \bra{f(\beta)} &=& \bra{0(\beta)} b_{{k}^\prime} (\beta)b_{{p}^\prime}(\beta)  .
\end{eqnarray}
Here, we consider $\hat{S}^{(2)}$ as the second-order term of the scattering matrix, given by
\begin{eqnarray}
    \hat{S}^{(2)}=\frac{(-i)^2}{2!}\int d^4x \, d^4y\mathcal{T}\left[\hat{\mathcal{L}}_I(x)\hat{\mathcal{L}}_I(y)\right],
\end{eqnarray}
where  $\hat{\mathcal{L}}_I=\mathcal{L}_I-\tilde{\mathcal{L}}_I$ is the interaction Lagrangian in the doubled space, and is given by
\begin{eqnarray}
    \hat{\mathcal{L}}_I=-g\Bar{\psi}\phi\psi+g\tilde{\bar{\psi}}\tilde{\phi}\tilde{\psi}.
\end{eqnarray}
While the non-tilde terms represent the conventional Hilbert space, the tilde terms represent the dual space arising from the doubling of this representation, resulting in parameters analogous in structure to those in the standard space. This construction is necessary to introduce thermal effects via TFD formalism \cite{temp000, khannatfd}. The physical quantities are described by the non-tilde operators.
Substituting the interaction Lagrangian, the transition amplitude becomes
\begin{equation}
\label{5}
    i\mathcal{M}_{fi}^{(\beta)} (2\pi)^4 \delta (\Sigma p) = \frac{(-ig)^2}{2!}\int d^4x \, d^4y \bra{f(\beta)} \mathcal{T}\left[ \Bar{\psi}_x\phi _x\psi _x \Bar{\psi}_y \phi _y\psi _y\right] \ket{i(\beta)},
\end{equation}
where the delta function guarantees that momentum conservation is maintained. Here, only the non-tilde part is considered; however, a similar result holds for the tilde part. The only non-vanishing contributions are from terms where the contractions occur between the field operators and the creation and annihilation operators of the initial and final particles. 
Solving Eq. (\ref{5})  we obtain
\begin{eqnarray}
    \label{}
    \nonumber i\mathcal{M}^{(\beta)}(2\pi)^4 \delta (\Sigma p) &=&  g^2F(\beta)\int d^4x \frac{d^4q}{(2\pi)^4} \, d^4y \Bar{u}_{{p}^\prime}\, u_p \Bar{u}_{{k}^\prime} u_k e^{i(p^\prime-p)x} e^{i(k^\prime-k)y} D_F^{11}(q) e^{-iq(x-y)} \\
    &-& g^2F(\beta)\int d^4x \frac{d^4q}{(2\pi)^4} \, d^4y \Bar{u}_{{p}^\prime}\, u_k \Bar{u}_{{k}^\prime} u_p e^{i(p^\prime-k)x} e^{i(k^\prime-p)y} D_F^{11}(q) e^{-iq(x-y)} \, ,
\end{eqnarray}
 where $F(\beta)=U_{p^\prime}(\beta)U_{p}(\beta)U_{k^\prime}(\beta)U_{k}(\beta)$, $D^{11}_F(q)$ is the $(a,b)=(1,1)$ component of the doubled-space bosonic Feynman propagator on the momentum space. Assuming that total momentum is conserved, the thermal transition amplitude can be written as
\begin{eqnarray}
    i\mathcal{M}^{(\beta)}&=&g^2F(\beta)\left[\bar{u}_{p^\prime}u_pD_F^{11}(p^\prime-p)\bar{u}_{k^\prime}u_k-\bar{u}_{p^\prime}u_kD_F^{11}(p^\prime-k)\bar{u}_{k^\prime}u_p\right].\label{eq08}
\end{eqnarray}
This result corresponds to the sum of two Feynman diagrams depicted in Figure \ref{fig1}. It is often denoted as $i\mathcal{M}^{(\beta)}=i\mathcal{M}_{a}^{(\beta)}+i\mathcal{M}_{b}^{(\beta)}$, where each index refers to diagrams (a) and (b) in the Figure \ref{fig1}.

In the next subsection, we will calculate { the differential cross section for this scattering process.}

{ \subsection{Yukawa scattering differential cross-section}\label{seccross}}

In order to obtain the cross section for the Yukawa scattering process, let's start by rewriting Eq. (\ref{eq08}) in a more convenient way, such that,\begin{eqnarray}
    i\mathcal{M}^{(\beta)}&=& i\mathcal{M}_{a}^{(\beta)}+i\mathcal{M}_{b}^{(\beta)}.
\end{eqnarray}
In such a way that the helicity average of the probability density is given by
\begin{eqnarray}
    \langle|\mathcal{M}^{(\beta)}|^2\rangle =\frac{1}{4}\sum_{s}|\mathcal{M}^{(\beta)}|^2=\bigl\langle|\mathcal{M}^{(\beta)}_{a}|^2\bigr\rangle +\bigl\langle|\mathcal{M}^{(\beta)}_{b}|^2\bigr\rangle +\bigl\langle2\Re{\mathcal{M}_{a}^{(\beta)\dagger}\mathcal{M}_{b}^{(\beta)}}\bigr\rangle,\label{Mvalue}
\end{eqnarray}
where $\langle|\mathcal{M}_{a}^{(\beta)}|^2\rangle$ and $\langle|\mathcal{M}_{b}^{(\beta)}|^2\rangle$ represent the two Feynman diagrams in Figure \ref{fig1}, and the third is an interference term between them.

Using the Mandelstam variables defined as
\begin{eqnarray}
    p\cdot k=s/2-m^2,\quad\quad p\cdot p^\prime=m^2-t/2,\quad\quad p\cdot k^\prime=m^2-u/2,
\end{eqnarray}
we can write the average values of the scattering amplitudes up to the second order of the LV-factor as
\small{\begin{eqnarray}
\bigl\langle|\mathcal{M}^{(\beta)}_{a}|^2\bigr\rangle&=&g^4F^2(\beta)(t-4m^2)^2\biggl\{\frac{1}{(t-m_\phi^2)^2}\bigg[1-\frac{2f_b(q_1)}{(t-m_\phi^2)}+\frac{3f_b^2(q_1)}{(t-m_\phi^2)^2}\bigg]\nonumber\\
&+&\frac{(2\pi)^2}{(e^{\beta E_{CM}}-1)^2}\biggl[\delta^2(q_1^2-m_\phi^2)-2f_b(q_1)\delta(q_1^2-m_\phi^2)\delta^\prime(q_1^2-m_\phi^2)+f_b^2(q_1)\delta^{\prime2}(q_1^2-m_\phi^2)\biggr]\biggr\},\label{eqM11}
    \end{eqnarray}}
where $(p^\prime-p)^\mu=q_1^\mu$ and $f_b(q)=b_{\mu\nu}q^{\mu}q^{\nu}$. Similarly, for the other terms,
\begin{eqnarray}
    \bigl\langle|\mathcal{M}^{(\beta)}_{b}|^2\bigr\rangle&=&g^4F^2(\beta)(u-4m^2)^2\biggl\{\frac{1}{(u-m_\phi^2)^2}\bigg[1-\frac{2f_b(q_2)}{(u-m_\phi^2)}+\frac{3f_b^2(q_2)}{(u-m_\phi^2)^2}\bigg]\nonumber\\&+&\frac{(2\pi)^2}{(e^{\beta E_{CM}}-1)^2}\biggl[\delta^2(q_2^2-m_\phi^2)-2f_b(q_2)\delta(q_2^2-m_\phi^2)\delta^\prime(q_2^2-m_\phi^2)+f_b^2(q_2)\delta^{\prime2}(q_2^2-m_\phi^2)\biggr]\biggr\},\label{eqM12}
    \end{eqnarray}
with $(p^\prime-k)^\mu=q_2^\mu$, and
\begin{eqnarray}
    \bigl\langle2\Re{\mathcal{M}_{a}^{(\beta)\dagger}\mathcal{M}_{b}^{(\beta)}}\bigr\rangle&=&-\frac{g^4F^2(\beta)}{2}\frac{[16m^4+8m^2(s-t-u)-s^2+t^2+u^2]}{(t-m^2_{\phi})(u-m^2_{\phi})}\nonumber\\&\times&\biggl\{\bigg[1-\frac{f_b(q_1)}{(t-m_\phi^2)}-\frac{f_b(q_2)}{(u-m_\phi^2)}+\frac{f_b^2(q_1)}{(t-m_\phi^2)^2}+\frac{f^2_b(q_2)}{(u-m_\phi^2)^2}+\frac{f_b(q_1)f_b(q_2)}{(u-m_\phi^2)(t-m_\phi^2)}\bigg]\nonumber\\
    &+&\frac{(2\pi)^2(t-m_{\phi})(u-m_{\phi})}{(e^{\beta E_{CM}}-1)^2}\biggl[\delta(q_1^2-m_{\phi}^2)\delta(q_2^2-m_{\phi}^2)-f_b(q_1)\delta^\prime(q_1^2-m_{\phi}^2)\delta(q_2^2-m_{\phi}^2)\nonumber\\&-&f_b(q_2)\delta(q_1^2-m_{\phi}^2)\delta^\prime(q_2^2-m_{\phi}^2)+f_b(q_1)f_b(q_2)\delta^\prime(q_1^2-m_{\phi}^2)\delta^\prime(q_2^2-m_{\phi}^2)\biggr]\biggr\}.\label{eqM13}
\end{eqnarray}

{ Using the center-of-mass frame of reference, the momenta can be expressed as follows
\begin{eqnarray}p=(E,\vec{p}_i);\quad\quad k=(E,-\vec{p}_i);\quad\quad p^\prime=(E,\vec{p}_f);\quad\quad k^\prime=(E,-\vec{p}_f),\end{eqnarray}
in such a way that the Mandelstam variables are given by
\begin{eqnarray}s=(2E)^2;\quad\quad t=2(m^2-E^2)(1+\cos{\theta});\quad\quad u=2(m^2-E^2)(1-\cos{\theta}).\end{eqnarray}
With $\theta$ being the angle between $\vec{p}_i$ and $\vec{p}_f$. Additionally, for a scattering process where all four masses are equal, the expression for the differential cross section is given by \cite{peskin}
\begin{eqnarray}\left(\frac{d\sigma}{d\Omega}\right)=\frac{|\mathcal{M}|^2}{64\pi^2 s},\label{sigma}
\end{eqnarray}
where the transition amplitude is calculated from Eqs. (\ref{eqM11})-(\ref{eqM13}).

It is worth mentioning that a power series expansion for very small $b_{ij}$ has been used. For the propagator, the expansion described in Eq. (\ref{eq09}) was applied. By combining these results into Eq. (\ref{Mvalue}) and substituting them into Eq. (\ref{sigma}), we obtain the differential cross section at finite temperature with Lorentz violation, as follows
\begin{eqnarray}\left(\frac{d\sigma}{d\Omega}\right)&=&\frac{g^4F^2(\beta)}{256\pi^2 E^2}\Bigg\{\biggl[\frac{\Gamma^{+}-4m^2}{\Gamma^{+}-m_\phi^2}\biggr]^2\left[1-\frac{2f_b(q_1)}{\Gamma^{+}-m_\phi^2}+3\left(\frac{f_b(q_1)}{\Gamma^{+}-m_\phi^2}\right)^2\right]\nonumber\\
&+&\biggl[\frac{\Gamma^{-}-4m^2}{\Gamma^{-}-m_\phi^2}\biggr]^2\left[1-\frac{2f_b(q_2)}{\Gamma^{-}-m_\phi^2}+3\left(\frac{f_b(q_2)}{\Gamma^{-}-m_\phi^2}\right)^2\right]+\frac{-16E^2m^2+4\Pi_{-}^2\sin^2\theta}{4\Pi_{-}^2\sin^2{\theta}+4\Pi_{-}m_\phi^2+m_\phi^4}\Bigg\}\nonumber\\
&+&\frac{g^4F^2(\beta)}{64E^2(e^{\beta E_{CM}}-1)^2}\biggl\{4\left[\Pi_{+}+\Pi_{-}\cos{\theta}\right]^2\biggl[\delta\biggl(\Gamma^{-}+m_\phi^2\biggr)-f_b(q_1)\delta^\prime\biggl(\Gamma^{-}+m_\phi^2\biggr)\biggr]^2\nonumber\\
&+&4\left[\Pi_{+}-\Pi_{-}\cos{\theta}\right]^2\biggl[\delta\biggl(\Gamma^{+}+m_\phi^2\biggr)-\delta^\prime\biggl(\Gamma^{+}+m_\phi^2\biggr)\biggr]^2-2\biggl[\Pi_{-}^2(1-\cos 2\theta)-8E^2m^2\biggr]\nonumber\\
&\times&\biggl[\delta\biggl(\Gamma^{-}+m_\phi^2\biggr)-f_b(q_1)\delta^\prime\biggl(\Gamma^{-}+m_\phi^2\biggr)\biggr]\biggl[\delta\biggl(\Gamma^{+}+m_\phi^2\biggr)-\delta^\prime\biggl(\Gamma^{+}+m_\phi^2\biggr)\biggr]\biggr\}\label{eq22}
\end{eqnarray}
 where have been used that
\begin{eqnarray}
\Gamma^{\pm}&=&2(m^2-E^2)(1\pm\cos{\theta}),\\
\Pi_{\pm}&=&E^2\pm m^2
\end{eqnarray}
and function $f_b(q_{1/2})$ is defined as
\begin{eqnarray}f_b(q_{1/2})=(E^2-m^2)\biggl[b_{11}\sin^2{\theta}+(b_{13}+b_{31})\sin{\theta}(\cos{\theta}\mp1)+b_{33}(\cos{\theta}\mp1)^2\biggr]\end{eqnarray}
where the upper (lower) sign represents the $q_1$ ($q_2$) variable. In addition to that
\begin{eqnarray}F(\beta)=\frac{1}{4}\biggl[\coth{\biggl(\frac{\beta E}{2}\biggr)}+1\biggr]^2.\label{eq21}
\end{eqnarray}

In this approach, the violation manifests as a background field in the kinetic component of the Lagrangian, which modifies the propagator. Additionally, temperature introduces an overall multiplicative factor in the final result $F(\beta)$, which in this frame of reference, is given by Eq. (\ref{eq21}). It is important to note that similar functions are obtained in \cite{scatter2, cabral2023violation, scatter3}, which differ from one another due to the interacting particles and the chosen reference frame.
Notice that the temperature dependence acts in such a way that it becomes the dominant factor in the cross-section as the temperature increases. Additionally, all the results presented involve a product of delta functions. This is common in the study of real-time thermal formalism and is not problematic, as regularization methods are available to handle these delta functions \cite{regdelta}.}

Now, let's consider two limiting cases where the tensor $b_{\mu \nu}$ is classified as time-like and space-like.

\subsubsection{Time-like case}

Note that, by momentum transfer, what we have are two fermions exchanging momentum (in the $z-$direction) and energy, through a virtual boson, in such a way that the energy and the momentum $p_z$ are conserved. So under these conditions, there are some features of the Sec. \ref{sec1} of a free scalar particle that we need to revisit.

By time-like, we define the symmetric traceless tensor $b_{\mu \nu}$ in the form given by
\begin{eqnarray}
   b_{\mu\nu}= \begin{pmatrix}
         0 & b_{01} & b_{02} & b_{03} \\ b_{01} &0&0&0\\
          b_{02} &0&0&0\\
          b_{03} &0&0&0
    \end{pmatrix},
\end{eqnarray}
which implies the following form for the stress-energy tensor in the free theory (\ref{eq26})
\begin{eqnarray}
    T^{\alpha\beta}=T^{\alpha\beta}_{\text{LI}}+\biggl(\delta^{\alpha}_0b^{0\mu}+\delta^{\alpha}_3b^{3\mu}\biggr)\partial_\mu\phi\partial^\beta\phi-g^{\alpha\beta}b^{0\mu}\partial_0\phi\partial_\mu\phi,
\end{eqnarray}
where $T^{\alpha\beta}_{\text{LI}}$ represents the Lorentz-invariant component of the energy-momentum tensor. This tells us that $T^{00} = T^{00}_{\text{LI}}$ and $T^{33} = T^{33}_{\text{LI}}$, implying that there are no contributions from Lorentz violation to the energy and pressure components of the stress-energy tensor.

Regarding Yukawa scattering, we simply need to substitute these { LV functions into Eq. (\ref{eq22})
\begin{eqnarray}
    f_b(q_1)=f_b(q_2)=0,
\end{eqnarray}
which makes the cross section explicitly independent of the LV tensor.}

\subsubsection{Space-like case}

We define the space-like regime by the tensor $b_{\mu \nu}$ in the form given by
\begin{eqnarray}
 b_{\mu\nu}=   \begin{pmatrix}
          0&0&0&0 \\ 
          0&b_{11}&b_{12}&b_{13}\\
          0&b_{12}&b_{22}&b_{23}\\
          0&b_{13}&b_{23}&b_{33}
    \end{pmatrix},
\end{eqnarray}
which leads to
\begin{eqnarray}
T^{\alpha\beta}=T^{\alpha\beta}_{\text{LI}}-\delta^\alpha_3b_{33}\partial^3\phi\partial^\beta\phi-\frac{1}{2}g^{\alpha\beta}b_{33}\partial^3\phi\partial^3\phi.
\end{eqnarray}
Therefore, the contributions to the energy and pressure are
\begin{eqnarray}
T^{00}=T^{00}_{\text{LI}}+\frac{1}{2}b_{33}k_z^2\phi^2\quad\quad\text{and}\quad\quad T^{33}=T^{33}_{\text{LI}}+\frac{1}{2}b_{33}k_z^2\phi^2,
\end{eqnarray}
implying that the violation directly affects the stress-energy tensor.

Additionally, the { LV functions that modify the cross-section are given by
\begin{eqnarray}
    f_b(q_{1/2})=(E^2-m^2)\biggl[b_{11}\sin^2{\theta}+2b_{13}\sin{\theta}(\cos{\theta}\mp1)+b_{33}(\cos{\theta}\mp1)^2\biggr].
\end{eqnarray}}
Note that, although the background tensor is traceless, we do not necessarily have $b_{33}=0$. What we have, in fact, is $g^{\mu\nu}b_{\mu\nu}=0$,  which implies $b_{11}+b_{22}+b_{33}=0$.

The discussions here for the time-like and space-like approaches demonstrate that all components of the violation modify the cross-section in specific ways. However, only the space-like components affect the stress-energy tensor, altering the distribution of energy and momentum in space-time.

Other limits of this scattering process can also be analysed, particularly the Lorentz-violating cross-section at zero temperature and the Lorentz-invariant version of this quantity, as detailed in the following subsection.

\subsubsection{Special limits}

At zero temperature, taking the limit $\beta \to \infty$ in Eq. (\ref{eq22}) implies that $F(\beta) \to 1$, and the {differential cross section  becomes
 \begin{eqnarray}
\left(\frac{d\sigma}{d\Omega}\right)_{LV}&=&\frac{g^4}{256\pi^2 E^2}\Biggl\{\biggl[\frac{\Gamma^{+}-4m^2}{\Gamma^{+}-m_\phi^2}\biggr]^2\left[1-\frac{2f_b(q_1)}{\Gamma^{+}-m_\phi^2}+3\left(\frac{f_b(q_1)}{\Gamma^{+}-m_\phi^2}\right)^2\right]\nonumber\\
&+&\biggl[\frac{\Gamma^{-}-4m^2}{\Gamma^{-}-m_\phi^2}\biggr]^2\left[1-\frac{2f_b(q_2)}{\Gamma^{-}-m_\phi^2}+3\left(\frac{f_b(q_2)}{\Gamma^{-}-m_\phi^2}\right)^2\right]+\left.\frac{-16E^2m^2+4\Pi_{-}^2\sin^2\theta}{4\Pi_{-}^2\sin^2{\theta}+4\Pi_{-}m_\phi^2+m_\phi^4}\right\}\;.\end{eqnarray}

Thus, there are first- and second-order corrections in the $b_{\mu \nu}$ tensor through the { LV functions $f_b(q_1)$ and $f_b(q_2)$}. Moreover, in the limit where the process is Lorentz invariant ($b_{ij} \to 0$) and there are no thermal effects ($T \to 0$), we obtain
\begin{eqnarray}
   \left(\frac{d\sigma}{d\Omega}\right)_{0}&=&\frac{g^4}{256\pi^2 E^2}\biggl\{\biggl[\frac{\Gamma^{+}-4m^2}{\Gamma^{+}-m_\phi^2}\biggr]^2+\biggl[\frac{\Gamma^{-}-4m^2}{\Gamma^{-}-m_\phi^2}\biggr]^2+\frac{-16E^2m^2+4\Pi_{-}^2\sin^2\theta}{4\Pi_{-}^2\sin^2{\theta}+4\Pi_{-}m_\phi^2+m_\phi^4}\biggr\}.
\end{eqnarray}
Notice that the functional dependence of any cross-section is solely on the energy $E$. This illustrates that the particle being collided is on the target, and we can only control the initial particle beams with a fixed energy.

\section{Yukawa Potential}\label{secpotential}

Another important quantity in Yukawa theory is the potential, a non-relativistic quantity that describes the interaction between two fermions, as depicted in the Feynman diagrams of Figure \ref{fig1}, in a classical framework. In this section, we will demonstrate how thermal effects and Lorentz symmetry breaking \cite{altschul} can influence the Yukawa potential.

In the non-relativistic limit, the second Feynman diagram in Figure \ref{fig1} is purely relativistic and falls outside this context. Therefore, the reaction is determined exclusively by the first diagram. Writing it explicitly, we have
\begin{eqnarray}
    i\mathcal{M}_{a}^{(\beta)}&=&-ig^2F(\beta)\bar{u}_{p^\prime}u_p\bar{u}_{k^\prime}u_k\biggl\{\frac{1}{\eta^{\mu\nu}(p^\prime-p)_\mu(p^\prime-p)_\nu-m_{\phi}^2}\nonumber\\&+&\frac{2\pi i}{e^{\beta (E_{p}+E_{p^\prime})}-1}\delta[\eta^{\mu\nu}(p^\prime-p)_\mu(p^\prime-p)_\nu-m_\phi^2]\biggr\}.
\end{eqnarray}
In this approach, we can write the momenta as
\begin{eqnarray}
    p=(m,\vec{p});\quad\quad k=(m,\vec{k});\quad\quad p^\prime=(m,\vec{p}^\prime);\quad\quad k^\prime=(m,\vec{k}^\prime);\label{eq05}
\end{eqnarray}
in addition to
\begin{eqnarray}
    (p^\prime-p)^2\sim -|\vec{p}^\prime-\vec{p}|^2,\label{eq06}
\end{eqnarray}
and write that the thermal function $F(\beta)$, as well as the LV term, are given by
\begin{eqnarray}
    F_{NR}(\beta)=\frac{1}{4}\biggr[\coth{\left(\frac{\beta m}{2}\right)}+1\biggl]^2
\end{eqnarray}
and
\begin{eqnarray}
    \eta^{\mu\nu}(p^\prime-p)_\mu(p^\prime-p)_\nu\sim -|\vec{q}|^2+b_{ij}q_iq_j.
\end{eqnarray}
Here, we define $\vec{q}=\vec{p}^\prime-\vec{p}$ for brevity. Notice that in the non-relativistic approach, it does not make sense to separately consider the temporal and spatial components of Lorentz violation as we did for the scattering. The only contribution comes from the space-like components.

In this context, we can directly express the Yukawa potential as
\begin{eqnarray}
    V_T(r)=-g^2F_{NR}(\beta)\biggl\{\int\frac{d^3q}{(2\pi)^3}\frac{e^{i\vec{q}\vec{r}}}{\vec{q}^2-b_{ij}q_iq_j+m_\phi^2}-\frac{2\pi i}{e^{\beta E_q}-1}\int\frac{d^3q}{(2\pi)^3}\delta\biggl[\vec{q}^2-b_{ij}q_iq_j+m_\phi^2\biggr]e^{i\vec{q}\vec{r}}\biggr\},
\end{eqnarray}
with $E_q=E_p+E_{p^\prime}$. To solve this integral, note that $q$ is a Euclidean momentum that can be transformed into a non-Euclidean version, $\bar{q}$, through a coordinate transformation. This new coordinate system has a non-diagonal metric $G_{ij}=\delta_{ij}-b_{ij}$ (a very small transformation), such that $d^3q \to \frac{d^{3}\bar{q}}{\sqrt{|G|}}$, where $|G|=\det(G_{ij})$. Therefore,
\begin{eqnarray}
(\delta_{ij}-b_{ij})q_iq_j=G_{ij}q_iq_j\to\bar{q}\quad \mbox{and}\quad (\delta_{ij}+b_{ij})r_ir_j=G_{ij}^{-1}r_ir_j.
\end{eqnarray}
Hence, the position transforms inversely to the momentum, so the product $\vec{r}\vec{q} \to \vec{\bar{r}}\vec{\bar{q}}$ is preserved. In other words, we have
\begin{eqnarray}
    V_T(r)=-\frac{g^2F_{NR}(\beta)}{\sqrt{|G|}}\biggl\{\int\frac{d^3\bar{q}}{(2\pi)^3}\frac{e^{i\vec{\bar{r}}\vec{\bar{q}}}}{(\vec{\bar{q}})^2+m_\phi^2}-\frac{2\pi i}{e^{\beta E_q}-1}\int\frac{d^3\bar{q}}{(2\pi)^3}\delta\biggl[(\vec{\bar{q}})^2+m_\phi^2\biggr]e^{i\vec{\bar{r}}\vec{\bar{q}}}\biggr\}.\label{eq07}
\end{eqnarray}

Since the root of $q^2 + m_{\phi}^2$ lies on the complex axis and the integral is evaluated in real space, the second term of (\ref{eq07}) vanishes, leading to
\begin{eqnarray}
    V_T(r)&=&%-\frac{g^2F_{NR}(\beta)}{(2\pi)^2\sqrt{|G|}}\int_0^\infty\frac{q^2dq}{q^2+m_\phi^2}\left[\frac{e^{iq\bar{r}}-e^{-iq\bar{r}}}{iq\bar{r}}\right]=-\frac{g^2F_{NR}(\beta)}{(2\pi)^2(i\bar{r})\sqrt{|G|}}\int_{-\infty}^\infty\frac{qe^{iq\bar{r}}}{q^2+m_\phi^2}dq\nonumber\\ &=&
    -\frac{g^2F_{NR}(\beta)}{(2\pi)^2(i\bar{r})\sqrt{|G|}}\left[2\pi i\frac{im_\phi e^{-m_\phi \bar{r}}}{2im_\phi}\right]=-\frac{g^2F_{NR}(\beta)}{4\pi\bar{r}\sqrt{|G|}}e^{-m_\phi\bar{r}},
\end{eqnarray}
where the remaining integral was taken over the upper half of the complex plane, with the contour including the pole $q = i m_\phi$.

Upon a more careful analysis, we find
\begin{eqnarray}
    \bar{r}=\sqrt{\bar{r}\bar{r}}=\left[G_{ij}^{-1}r_ir_j\right]^{\frac{1}{2}}=\left[\left(\delta_{ij}+b_{ij}\right)r_ir_j\right]^{\frac{1}{2}}=\left[r^2+b_{ij}r_ir_j\right]^{\frac{1}{2}}=r\left[1+b_{ij}\hat{r}_i\hat{r}_j\right]^{\frac{1}{2}}.
\end{eqnarray}
Up to first order, we have $\det{G_{ij}} = 1 - b_{jj}$. Thus, expanding it, we get
\begin{eqnarray}
    \frac{1}{\bar{r}\sqrt{|G|}}e^{-m_\phi\bar{r}}&\sim&%\frac{1}{r(1+b_{ij}\hat{r}_i\hat{r}_j)^{\frac{1}{2}}}\frac{\exp{-m_{\phi}r(1+b_{ij}\hat{r}_i\hat{r}_j)^{\frac{1}{2}}}}{\sqrt{1-b_{jj}}}\nonumber\\&=&\frac{e^{-m_\phi r}}{r}\biggl[\biggl(1+\frac{1}{2}b_{jj}+\cdots\biggr)\biggl(1-\frac{1}{2}b_{ij}\hat{r}_i\hat{r}_j+\cdots\biggr)\exp{-\frac{1}{2}m_{\phi}rb_{ij}\hat{r}_i\hat{r}_j+\cdots}\biggr]\nonumber\\ &\sim&
    \frac{e^{-m_\phi r}}{r}\biggl[1+\frac{1}{2}b_{jj}-\frac{1}{2}(1+m_\phi r)b_{ij}\hat{r}_i\hat{r}_j\biggr].
\end{eqnarray}
Therefore, the Yukawa potential in the context of Lorentz violation, with thermal contribution,  becomes
\begin{eqnarray}
    V_T(r)
    &=&-\frac{g^2}{16\pi}\biggr[\coth{\left(\frac{\beta m}{2}\right)}+1\biggl]^2\frac{e^{-m_\phi r}}{r}\biggl[1+\frac{1}{2}b_{jj}-\frac{1}{2}(1+m_\phi r)b_{ij}\hat{r}_i\hat{r}_j\biggr],\label{eq19}
\end{eqnarray}
where, taking $\beta \to \infty$ we recover the result achieved by \cite{altschul} and, in addition, at the limit $b_{ij} \to 0$, we recover the fundamental Yukawa potential as given in \cite{peskin}.

To analyze the thermal effects, let us assume $b_{ij} = 0$ for the moment, neglecting the Lorentz symmetry breaking. Then
\begin{eqnarray}
    V_T(r)=-\frac{g^2 }{16\pi}\left[\coth{\left(\frac{m}{2k_BT}\right)}+1\right]^2\frac{e^{-m_{\phi}r}}{r}.\label{eq29}
\end{eqnarray}
This implies that the potential exhibits a significant enhancement due to temperature, as shown in Figure \ref{fig3}. Note that the negative sign indicates an attractive interaction, and as $T$ increases, $|V_T|$ also increases. Therefore, the interaction becomes stronger in hotter scenarios.
\begin{figure}[ht]
    \centering
\includegraphics[width=0.6\linewidth]{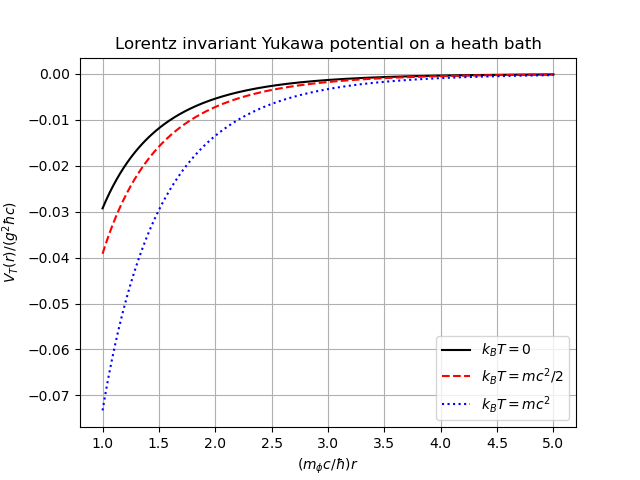}
    \caption{The thermal Yukawa potential given by Eq. (\ref{eq29}) as a function of the distance $r$ between the particles for temperatures $T=0$ (solid black line), $k_B T = mc^2 / 2$ (red dashed line), and $k_B T = mc^2$ (blue dotted line).}
    \label{fig3}
\end{figure}

Now, to focus solely on the symmetry-breaking contribution, we set $T=0$. Without loss of generality, assuming the LV coefficient is a traceless symmetric Lorentz tensor, Eq. (\ref{eq19}) becomes
\begin{eqnarray}
    V_{LV}(r)=-\frac{g^2}{4\pi}\frac{e^{-m_{\phi}r}}{r}\biggl[1+\frac{1}{2}(b_{x}+b_{y}+b_{z})-\frac{1}{2r^2}\biggl(1+{m_\phi}r\biggr)\biggl(b_{x}x^2+b_{y}y^2+b_{z}z^2\nonumber\\+2b_{xy}xy+2b_{xz}xz+2b_{yz}yz\biggr)\biggr].
\end{eqnarray}
It is evident that the presence of the LV term breaks rotational invariance. The function, which previously had no roots, now becomes zero in specific spatial regions, passing through the points
\begin{eqnarray}
    \biggl(\pm \frac{2+(b_y+b_z)}{b_x m_\phi},0,0\biggr);\quad\quad \biggl(0,\pm \frac{2+(b_x+b_z)}{b_y m_\phi},0\biggr);\quad\quad \biggl(0,0,\pm \frac{2+(b_x+b_y)}{b_z m_\phi}\biggr).
\end{eqnarray}

Additionally, to explore the possibility, let us assume $b_{ij} = \lambda$, where $\lambda$ is a very small number. Therefore, with the correct units, we obtain
\begin{eqnarray}
    V_{LV}(r)=-\frac{g^2}{4\pi}\frac{e^{-m_{\phi}r}}{r}\left[1+\frac{3\lambda}{2}-\frac{1}{2}\biggl(1+m_\phi r\biggr)\lambda\frac{(x+y+z)^2}{r^2}\right].\label{eq27}
\end{eqnarray}
This function is shown in Figure \ref{fig4}, from the perspective of the $z=0$ plane.
\begin{figure}[ht]
    \centering
    \subfigure[]{\includegraphics[scale=0.4]{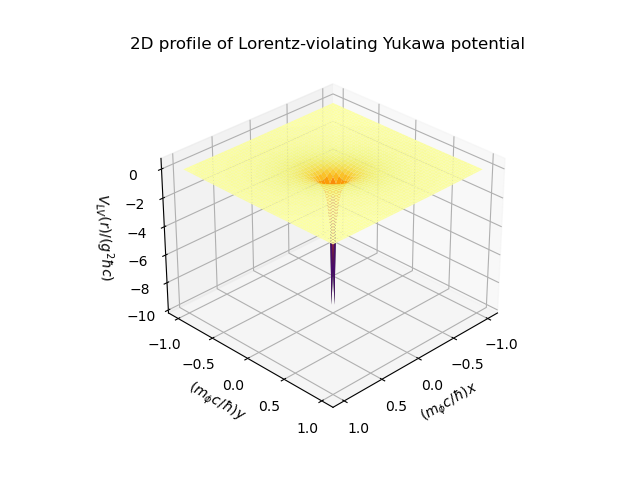}}
        \subfigure[]{\includegraphics[scale=0.4]{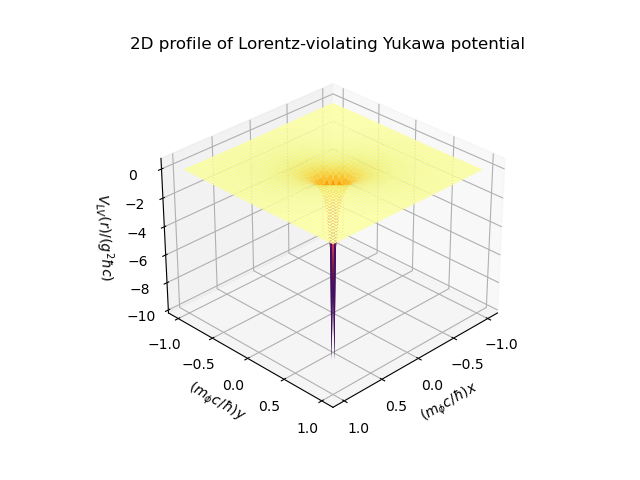}}
    \subfigure[]{\includegraphics[scale=0.4]{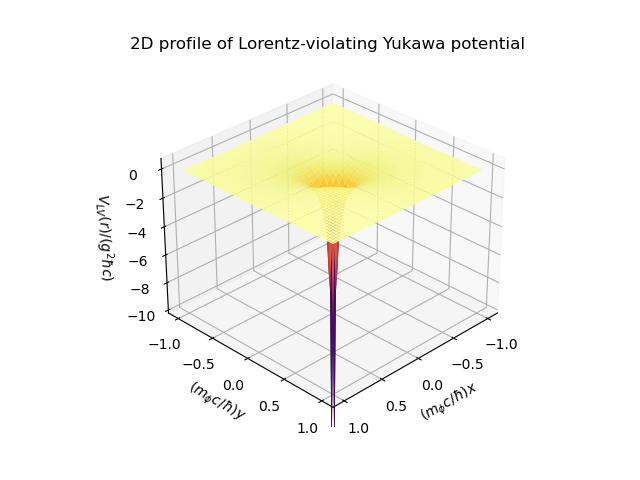}}
    \subfigure[]{\includegraphics[scale=0.4]{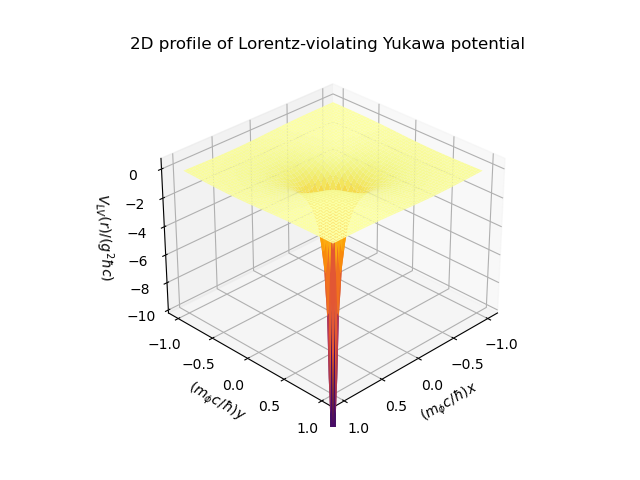}}
    \hfill
    \caption{The profile of the potential $V_{LV}(r)$ at zero temperature, as given by Eq. (\ref{eq27}), as a function of the distance $r$ in the $z=0$ plane, for $\lambda = 1/2$ (a), $\lambda = 1$ (b), $\lambda = 2$ (c), and $\lambda = 5$ (d). }
    \label{fig4}
\end{figure}

It is evident that the presence of Lorentz violation, even at zero temperature, significantly increases the strength of the interaction. However, this is not the only effect. The factor in Eq. (\ref{eq27}) shows a substantial modification, breaking rotational symmetry and deforming the potential surface. This results in different values for different coordinates, effectively choosing preferential directions.

Since the range of the interaction is defined by $\Delta V = V\left(\frac{1}{m_{\phi}}\right)$, i.e., the distance at which the potential decays to $e^{-1}$ of its initial value, we can analyze the change in this range from
\begin{eqnarray}
    \frac{\Delta V_{LV}}{\Delta V}=1+\frac{3\lambda}{2}-\lambda(x_0+y_0+z_0)^2,
\end{eqnarray}
with $x = \frac{x_0}{m_{\phi}}$ and similarly for $y$ and $z$, this implies that even if the distance is the same, the range can be larger or smaller depending on the direction being analyzed and the LV coefficient. To simplify the discussion, the profile of the Yukawa potential in the $x$-direction, with $y = z = 0$, is shown in Figure \ref{fig5}.

If we reintroduce the temperature dependence from Eq. (\ref{eq29}) into Eq. (\ref{eq27}), we obtain
\begin{eqnarray}
    V_{LV}(r)=-\frac{g^2}{16\pi}\left[\coth{\left(\frac{m}{2k_BT}\right)}+1\right]^2\frac{e^{-m_{\phi}r}}{r}\left[1+\frac{3\lambda}{2}-\frac{1}{2}\biggl(1+m_\phi r\biggr)\lambda\frac{(x+y+z)^2}{r^2}\right].\label{eq30}
\end{eqnarray}
This function is plotted in Figure \ref{fig55}, which shows the behavior of the one-dimensional profile of the potential under the effects of Lorentz violation and temperature.
\begin{figure}[ht]
    \centering
    \subfigure[{The potential $V(r)$ at $T=0$ for $\lambda=0$ (black solid line), $\lambda=1$ (red dashed line) and $\lambda=5$ (blue dotted line).}]{\includegraphics[width=0.4\linewidth]{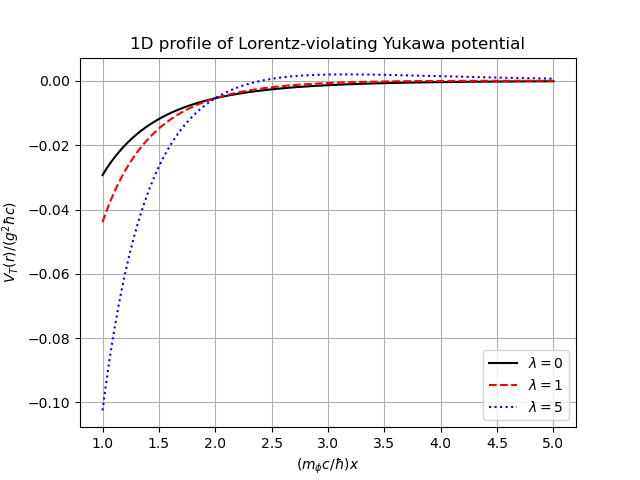} \label{fig5}}
    \subfigure[{The potential $V(r)$ at $\lambda=5$ for $k_BT=0$ (black solid line), $k_BT=mc^2/2$ (red dashed line) and $k_BT=mc^2$ (blue dotted line).}]{\includegraphics[width=0.4\linewidth]{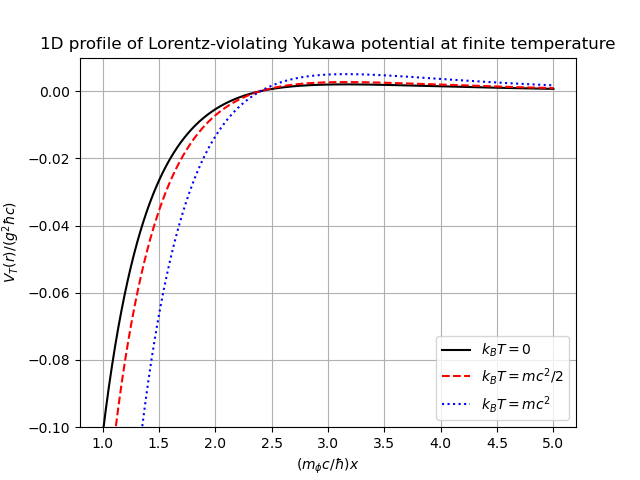}\label{fig55}}
    \caption{The 1-D profile of the potential $V(r)$ given by Eq. (\ref{eq30}) in function of $r=x$.}
   \end{figure}

The main observation is that temperature in Yukawa theory modifies both the cross-section and the potential. The modification of the potential is simpler than that of the cross-section: thermal corrections to the Yukawa potential primarily increase the strength of the interaction between particles. This enhancement is due exclusively to the function $F_{\text{NR}}(\beta)$, which depends on the temperature and the interacting lepton mass. Notably, heavier particles experience smaller changes due to the structure of the coth function.

Regarding Lorentz violation, we observe that it modifies the potential not only by increasing it but also by decreasing and potentially canceling it. These effects are closely dependent on the values of the LV parameters. From the one-dimensional profile described in Figure \ref{fig5}, we see that at very small distances, the interaction is stronger due to the presence of LV effects. As we increase the distance, the correlation between the particles weakens until the potential becomes repulsive, contrary to conventional theory. Notably, there is a point $R$ where $V(R) = 0$, indicating that at this distance, the particles behave as if they were moving freely. Additionally, there is a region where the potential exhibits a maximum, representing an unstable equilibrium similar to a harmonic interaction. These effects become more pronounced as the values of $b_{ij}$ increase.

The analysis of the Lorentz-violating thermal potential is straightforward, as it involves combining both features. In other words, Lorentz violation breaks all spatial symmetries, while the presence of thermal effects enhances the potential, as illustrated in Figure \ref{fig55}.

\section{Conclusion}\label{secconclusions}

In this work, we address the Yukawa theory by considering corrections from both Lorentz symmetry violation and the presence of a heat bath. The Lorentz-violating parameters are introduced into the problem through a kinetic term in the Lagrangian, which modifies the scalar boson propagator. We employed a real-time formalism for thermal field theory, deriving the propagator using the TFD formalism and incorporating temperature dependence through the doubling of the Hilbert space. In this context, the transition amplitude was derived, and the total cross section was calculated under these conditions. The results were analyzed for specific values of temperature and Lorentz-violating coefficients. Additionally, we presented the Yukawa potential, a non-relativistic quantity representing the interaction between two fermions, under the effects of temperature and Lorentz violation. We examined specific cases and discussed the resulting features. Finally, it was observed that considering Lorentz symmetry breakdown has several implications for changes in symmetries and physical states, while the presence of temperature is strongly linked to the strength of the interaction. Understanding the corrections due to both modifications is crucial for determining some characteristics of the early universe, as these factors are intrinsic to the primordial conditions.

\section*{Acknowledgments}

This work by A. F. S. is partially supported by National Council for Scientific and Technological
Development - CNPq project No. 312406/2023-1. D. S. C., L. A. S. E., J. C. R. S. and L. H. A. R. thank CAPES for financial support.

\section*{Data Availability Statement}

%No data are available because of the nature of the research. This publication is theoretical work that does not require supporting research data.
No Data associated in the manuscript.

\section*{Conflicts of Interest}

No conflict of interests in this paper.

%%%%%%%%%%%%%%%%%%%%%%%%%%%%%%%%%%%%%%%%%%%%%%%%%%%%%%%%%%%%%%%%%%%%%%%%%%%%%%%%%%%%%%%%%%%%%%%%%%%%%%%%%%%%%%%%%

\global\long\def\link#1#2{\href{http://eudml.org/#1}{#2}}
 \global\long\def\doi#1#2{\href{http://dx.doi.org/#1}{#2}}
 \global\long\def\arXiv#1#2{\href{http://arxiv.org/abs/#1}{arXiv:#1 [#2]}}
 \global\long\def\arXivOld#1{\href{http://arxiv.org/abs/#1}{arXiv:#1}}

%%%%%%%%%%%%%%%%%%%%%%%%%%%%%%%%%%%%%%%%%%%%%%%%%%%%%%%%%%%%%%%%%%%%%%%%%%%%%%%%%%%%%%%%%%%%%%%%%%%%%%%%%%%

\end{document}